\documentclass[final] {aipproc}
\usepackage{sidecap,graphicx,boxedminipage,epsfig,wrapfig,floatflt,shadow}

\layoutstyle{8x11double}
\pdfoutput=1
\usepackage{graphicx}

\begin{document}

\title{
Improving Students' Conceptual Understanding of Conductors and Insulators
}


\author{Joshua Bilak and Chandralekha Singh}{
  address={Department of Physics and Astronomy, University of Pittsburgh, Pittsburgh, PA, 15260}}

\begin{abstract}
We examine the difficulties that introductory physics students, undergraduate physics majors, and physics graduate
students have with concepts related to conductors and insulators covered in introductory
physics by giving written tests and interviewing a subset of students. We find that even
graduate students have serious difficulties with these concepts. We develop
tutorials related to these topics and evaluate their effectiveness by comparing the
performance on written pre-/post-tests and interviews
of students who received traditional instruction vs. those who learned using tutorials.

\end{abstract}

\maketitle

\section{Introduction}

Conductors and insulators are taught at increasing levels of mathematical sophistication starting from algebra- and 
calculus-based introductory physics (or high school physics) to graduate level physics courses~\cite{maloney,otero,gauss,chabay}. 
Considering the frequent appearance of these topics in electricity and magnetism (E$\&$M) courses at various levels,
one may assume that physics graduate students have a clear understanding of these concepts. 
Here, we investigate the conceptual difficulties that introductory students, undergraduate physics majors,
and physics graduate students have with concepts related to conductors and insulators covered at the introductory level.
We also develop tutorials and evaluate their effectiveness by comparing the performance of students who learned these
topics using traditional and tutorial instructions.

First, conceptual multiple-choice questions were developed and administered to introductory students and physics graduate 
students and interviews were conducted with some students.
We then developed tutorials and pre-/post-tests on these topics at the level of introductory physics.
We performed a controlled study and gave the pre-/post-tests in equivalent introductory physics courses before and after 
traditional or tutorial instruction. 
These tests require written responses often asking students for drawings and explanations. 
Interviews with introductory students and physics majors were conducted
using a ``think-aloud" protocol to gain further insight into their reasoning 
and the origins of their difficulties. In this protocol, students are asked to talk aloud as they work through the
material without interruption and they are asked for clarification of the points they had not otherwise made clear at the end.
Some physics majors worked on the tutorials while thinking aloud and were given the pre-/post-tests while other physics
majors
worked through the pre-/post-tests only without working on the tutorials.
In some interviews, introductory students and physics majors were shown relevant experimental setups, asked
to predict what should happen in certain situations and then asked to perform the experiment to check their prediction
and reconcile the differences between their original prediction and observation if they did not match.
We briefly discuss some findings.

\vspace*{-.20in}
\section{Results and Discussion}
\vspace*{-.11in}

We first discuss the performance of introductory students and graduate students on three conceptual
multiple-choice questions. Their performance shows that neither group has an adequate comprehension of these concepts. 
While 235 introductory students in three different classes
were administered these questions after instruction as part of a recitation quiz,
19 first year graduate students worked on them as part of a placement quiz for entry to a Jackson-level E$\&$M course.
Graduate students were told ahead of time that their placement quiz will be on undergraduate level E$\&$M.
Both graduate and introductory students made very similar mistakes and the commonality 
suggests that these concepts covered in introductory physics are indeed challenging. 
The questions described below deal with the charge distribution on a conductor with a charge outside,
the effect of grounding a conductor with a charge nearby, 
and the electric field inside a hollow insulator due to a charge outside:

(1) A small aluminum ball hanging from a thread is placed at the center of a tall metal cylinder and a positively charged plastic rod
is brought near the aluminum ball in such a way that the metal wall is in between the rod and the ball.
Which one of the following is the correct qualitative picture of the charge distribution in this situation?

\begin{center}
\includegraphics[width=3.1in]{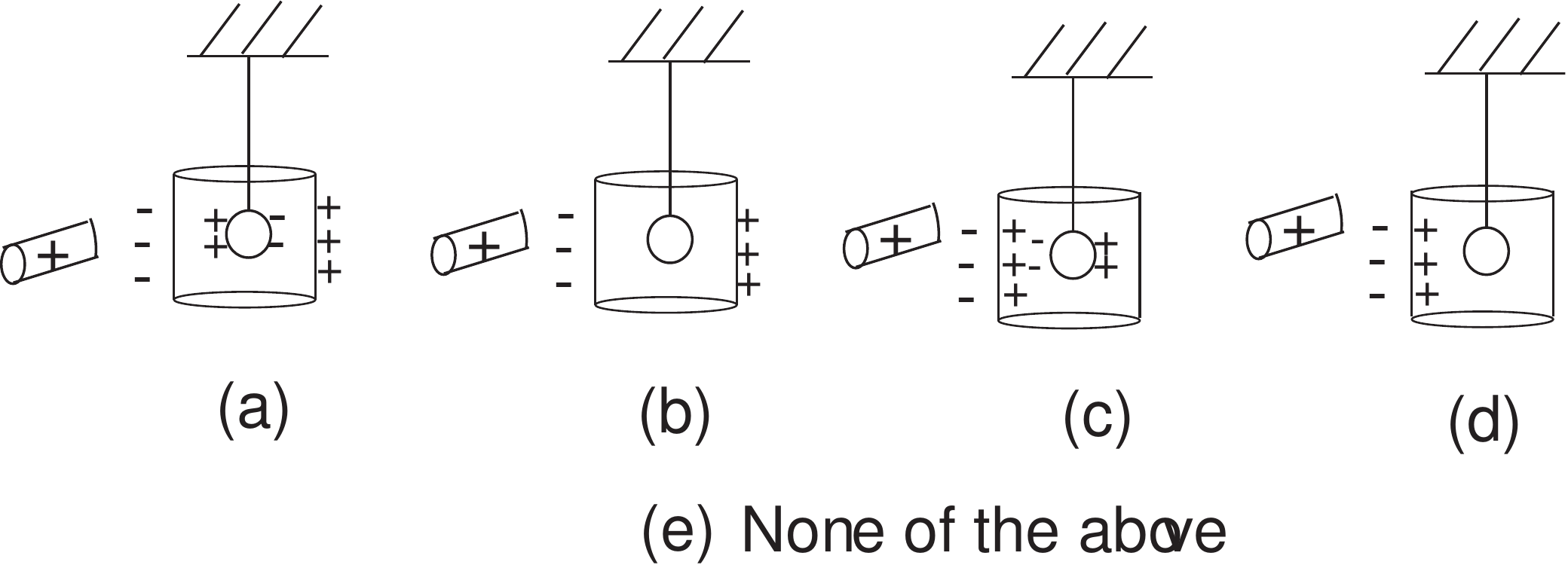}
\end{center}

(2) A positively charged plastic rod is taken close to (but not touching) a neutral aluminum ball hanging from a thread.
The aluminum ball is grounded by connecting it with a wire that touches the ground.
Which one of the following statements is true about the force between the rod and the ball?

\vspace*{-.3in}
\begin{center}
\includegraphics[width=2.74in]{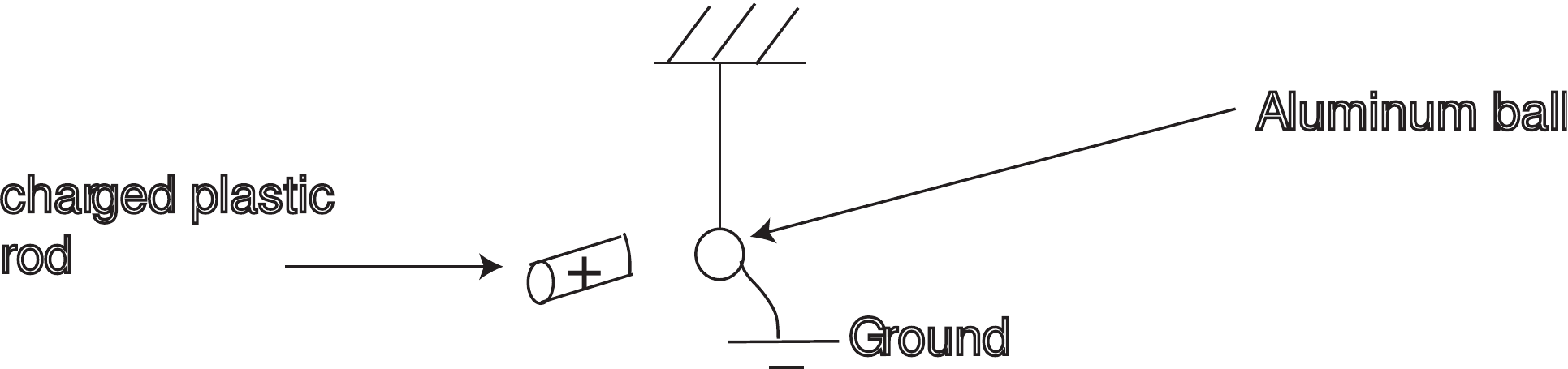}
\end{center}

\noindent
(a) The aluminum ball will not feel any force due to the charged rod.\\
(b) The aluminum ball will be attracted to the charged rod and the force of attraction is the same as that without the
grounding wire.\\
(c) The aluminum ball will be attracted to the charged rod and the force of attraction is more than that without the
grounding wire.\\
(d) The aluminum ball will be attracted to the charged rod and the force of attraction is less than that without the
grounding wire.\\
(e) The aluminum ball will be attracted or repelled 
depending upon the magnitude of the charge on the rod.\\

(3) A small aluminum ball hanging from a thread is placed at the center of a tall cylinder made with a good insulator 
(Styrofoam) and a positively charged plastic rod
is brought near the 
ball in such a way that the Styrofoam wall is in between the rod and the ball.
Which one of the following statements is true about this situation?

\begin{center}
\includegraphics[width=2.43in]{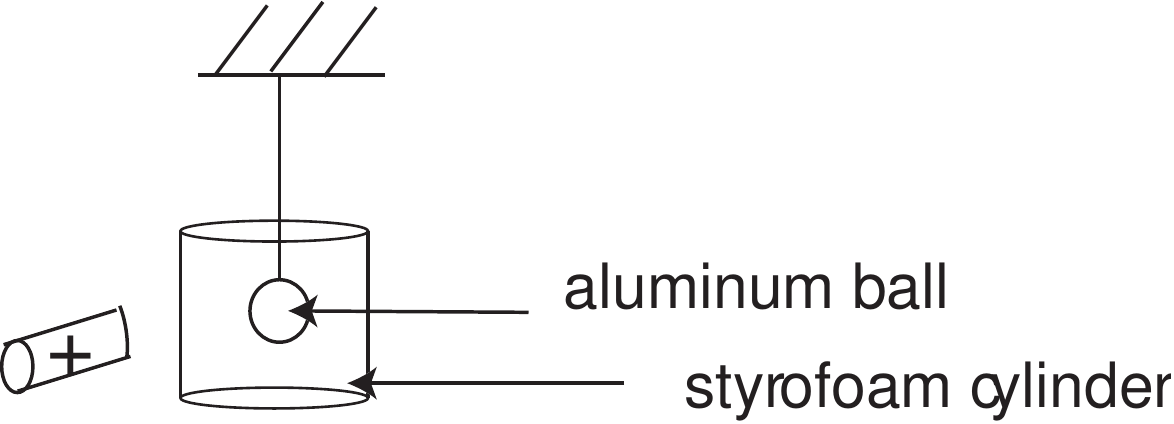}
\end{center}

\noindent
(a) The aluminum ball will not feel any force due to the charged rod.\\
(b) The aluminum ball will be attracted to the charged rod and the force of attraction is the same as that without the
Styrofoam cylinder.\\
(c) The aluminum ball will be attracted to the charged rod but the force of attraction is more than that without the
Styrofoam cylinder.\\
(d) The aluminum ball will be attracted to the charged rod and the force of attraction is less than that without the
Styrofoam cylinder.\\
(e) None of the above.

\hspace*{-1.5in}
\begin{table}[h]
\centering
\begin{tabular}[t]{|c|c|c|c|c|c|c|}
\hline
Q&	Group	&        a	&b	&c	&d	&e	\\[0.5 ex]
\hline
1 &	Introductory Students&	44&	{\it 23}&	26&	6&	1	 \\[0.5 ex]
&	Graduate Students&	32&	{\it 37}&	16&	5&	11	 \\[0.5 ex]
\hline
2 &	Introductory Students&	21&	15&	{\it 44}&	16&	1	 \\[0.5 ex]
&	Graduate Students&	16&	32&	{\it 53}&	0&	0	 \\[0.5 ex]
\hline
3 &	Introductory Students&	27&	1 &	12&	{\it 46} &	1	 \\[0.5 ex]
&	Graduate Students&	32&	26&	0&	{\it 42} &	0\\[0.5 ex]
\hline
\end{tabular}
\vspace{0.1in}
\caption{The average percentage of 235 introductory physics students and 19 physics graduate students who chose options (a)-(e) on
the three multiple-choice questions. The correct response for each question is italicized.
}
\label{junk2}
\end{table}

Table 1 shows that these questions were difficult and 
the performances of the introductory students and graduate students on each of these questions is not much different with neither
group performing better than $53\%$ on any of them.
While it is true that the introductory students had learned these concepts recently so they may be fresh in their minds, the questions
are sufficiently conceptual that we expected the graduate students to perform significantly better than they actually did.

Question (1) probes student's knowledge of the charge distribution 
via induction due to a charge outside the cavity. 
The correct response is option (b). The electric field inside the conductor is zero with no charge separation on the 
aluminum ball.
Approximately one-fourth of the introductory students and one-third of the graduate students provided the correct response.
The difficulties were common for both groups and $70\%$ of the introductory students and half of the graduate students 
believed that there will be a charge separation on the aluminum ball as well (options (a) and (c)). During individual interviews
with introductory students,
those who chose this option explained that the negative and positive charges on the wall of the hollow conductor will exert
forces on the electrons in the aluminum ball and induce a charge separation in the ball. When asked explicitly about the electric
field in the hollow region, these students often believed that there was a non-zero electric field inside. When asked explicitly
about the charges that produce the electric field in the hollow region, some students claimed that 
the charge on the plastic rod outside will also contribute to the field inside in addition to the induced charges.
Others claimed that it was only the charges
induced on the walls of the hollow region that produce the field inside because the electric field due to the
outside charge on the plastic rod was shielded by the conductor. 
Thus, although some of the students used the concept of shielding the inside of the conductor from the charges outside,
they did not realize that shielding implies that the net electric field inside the conductor is zero due to the cumulative effects
of the charges on the plastic rod and those induced on the outer walls of the conductor. Interviews suggest that some students chose
options (a) or (c) because they believed incorrectly that the situation given is similar to the one with a charge inside a
conductor cavity (perhaps because there was an aluminum ball in the cavity). Unlike the given situation, the charge inside a cavity can produce a field inside the cavity
in addition to inducing charges on the cavity walls.

In Question (2), if the aluminum ball with a positively charged rod nearby is grounded, free electrons come from the ground and neutralize the positive charge
on the aluminum ball so that the net attraction between the charged rod and the aluminum ball is more than when
there was no grounding wire (option (c)). Approximately half of the students from both groups did not provide the correct
response.
The incorrect responses suggest that one-third of the graduate students believed that the grounding wire will not affect the
force between the aluminum ball and the charged plastic rod whereas $16\%$ of them believed that a grounded object cannot be charged. 
The introductory students had similar difficulties. In a pretest given to introductory students before college instruction
about these concepts (in which they had to draw diagrams and explain their reasonings), students were asked a similar question
involving a grounded steel ball in the vicinity of a positively charged comb (see figure below). Some students who had
high school instruction in these concepts drew incorrect diagrams 
such as the following, 
in which protons travel from the 
ball to the ground and the electrons on the steel ball spread out uniformly on its surface while the comb is still nearby:

\begin{figure}[h!]
\includegraphics[width=2.03in]{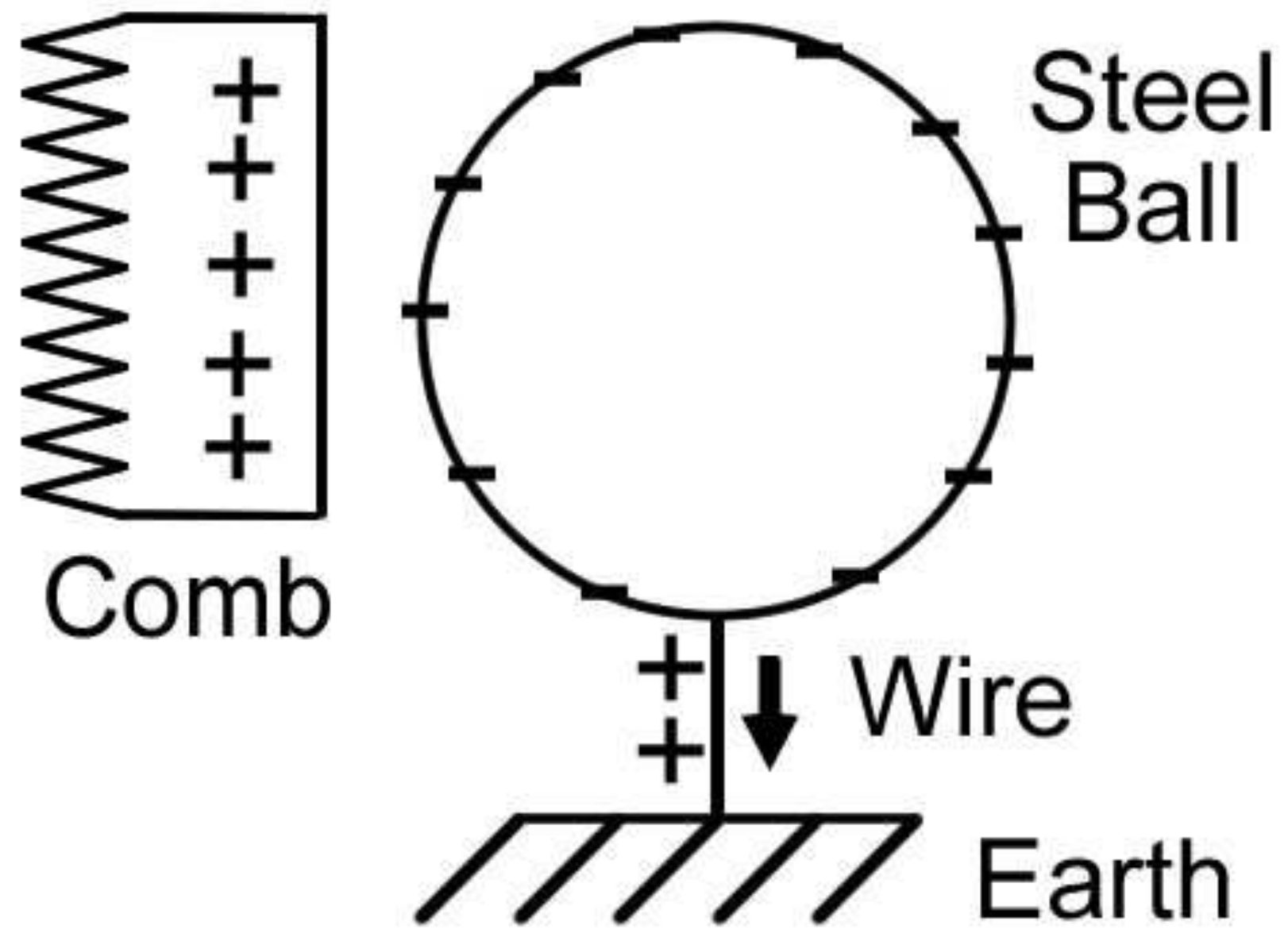}
\end{figure}

Written explanations and diagrams drawn on the pretest (before college instruction) and post-test (after college instruction)
given to introductory students and physics majors in the free-response format and
interviews with 
some of them suggest that students had
difficulty with the role of grounding and reasoning about the changes in the charge distribution due to grounding.
A major problem with this concept of grounding dealt with inaccurate characteristics attributed to the Earth. Since the Earth is a 
conductor, some students believed that it is negatively charged. This notion caused some
students to draw positive charges on the steel ball shifted downward due to the attraction from the negatively
charged Earth as in the following drawing by a student who answered the question while ``thinking-aloud":

\begin{figure}[h!]
\includegraphics[width=2.03in]{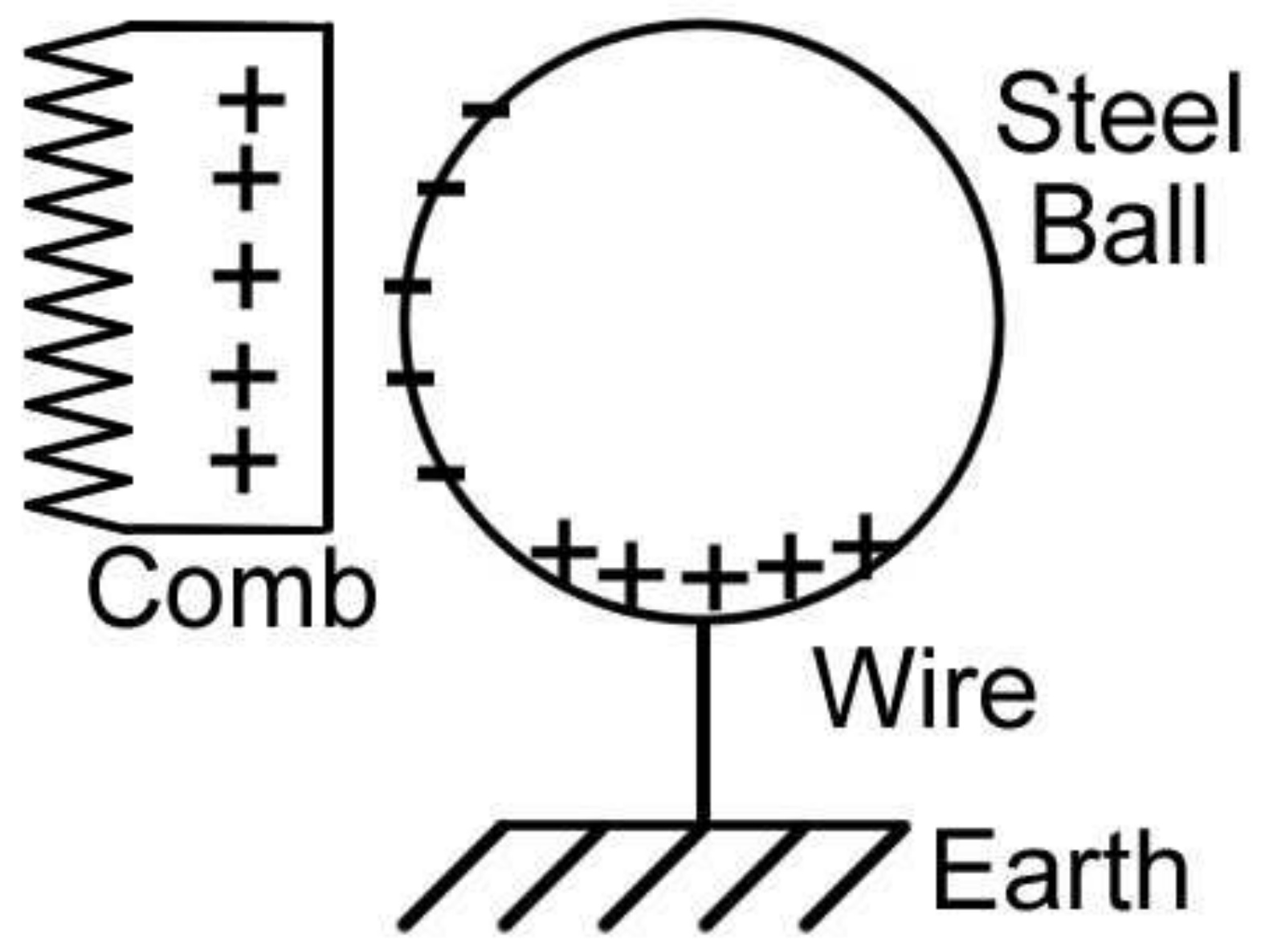}
\end{figure}

Question (3) asks how an aluminum ball, encased in an insulator (Styrofoam), 
will be affected by the presence of a positively charged rod outside. Since Styrofoam does not have
free electrons, the atoms in the Styrofoam will become polarized and the electric field inside the Styrofoam will be smaller than
that without the Styrofoam. Therefore, the force on the free electrons in the aluminum ball inside the Styrofoam (and hence
the separation of charges on the aluminum ball) will be smaller than the case without the Styrofoam. The force
of attraction between the plastic rod and the ball will be less than that without the Styrofoam (option (d)).
There is no significant difference between the performance of the introductory students and graduate students on this
question, with both groups obtaining less than $50\%$.

Surprisingly, approximately one-third of the graduate students invoked the notion of shielding and
believed that no force would be felt by the aluminum ball from the plastic rod outside. 
Introductory physics students and physics majors participating in the think-aloud protocol often
explained that an insulator will insulate the inside of the cylinder from the outside and prevent the electric
field due to the charges on the plastic rod from penetrating the walls and reaching the inside.
Written responses requiring explanations and interviews suggest that
many students took the word ``insulator" literally.
These students incorrectly claimed that a conducting wall around a cavity is not nearly as effective as an insulating
wall in preventing the electric field from the external sources from penetrating inside in analogy with an
insulating wall required to prevent heat transfer.
Asking students for the microscopic mechanism for zero field, i.e., asking them to
account explicitly for the charges whose electric fields will add up vectorially to make 
the net electric field zero everywhere inside the insulator turned out to be difficult for them.
Most students had not thought about why an electrical insulator will insulate the inside from the outside charges,
but believed in it intuitively. In some interviews,
after making the prediction that the inside of the Styrofoam is not influenced by the charges outside,
students were asked to
conduct the experiment, they observed that the aluminum ball inside the Styrofoam is attracted to the plastic rod
and the electric field inside the Styrofoam cannot be zero. However, this observation alone was not sufficient to help
them formulate correct reasoning about this situation.
We plan to evaluate student performance on these questions with the word insulator replaced with non-conductor.

These universal difficulties that persist even at the graduate level inspired us to develop tutorials and pre-/post-tests
about these concepts. The details of the tutorials that use a guided approach to helping students reason about these concepts
will be provided elsewhere. We performed a controlled study involving different 
introductory physics courses
to evaluate the effectiveness of the tutorials and gave the pre-/post-tests to students (before and after college
instruction) in classes using traditional and tutorial instructions. 
The pre-tests and post-tests required students to explain their reasoning 
and draw diagrams. We also interviewed some students. Incidentally, there was no significant difference in the performance
of students on the pre-test who had high school instruction in conductors and insulators vs. those who did not.
The graph in Figure 1 below shows that there was no significant improvement in introductory 
students' performance on the post-test compared to the pre-test when students were taught traditionally using
lecture alone (lec. group in the graph). The grading for these free-response questions was done using a scale where each
incorrect response was coded as 0, partially correct response as 1 and a correct response as 2. 
The pre-test scores for both groups (lec. and tut.) were combined because there was no significant difference.
The tutorial group performed significantly better on the post-test. 

\begin{figure}[h!]
\includegraphics[width=2.03in]{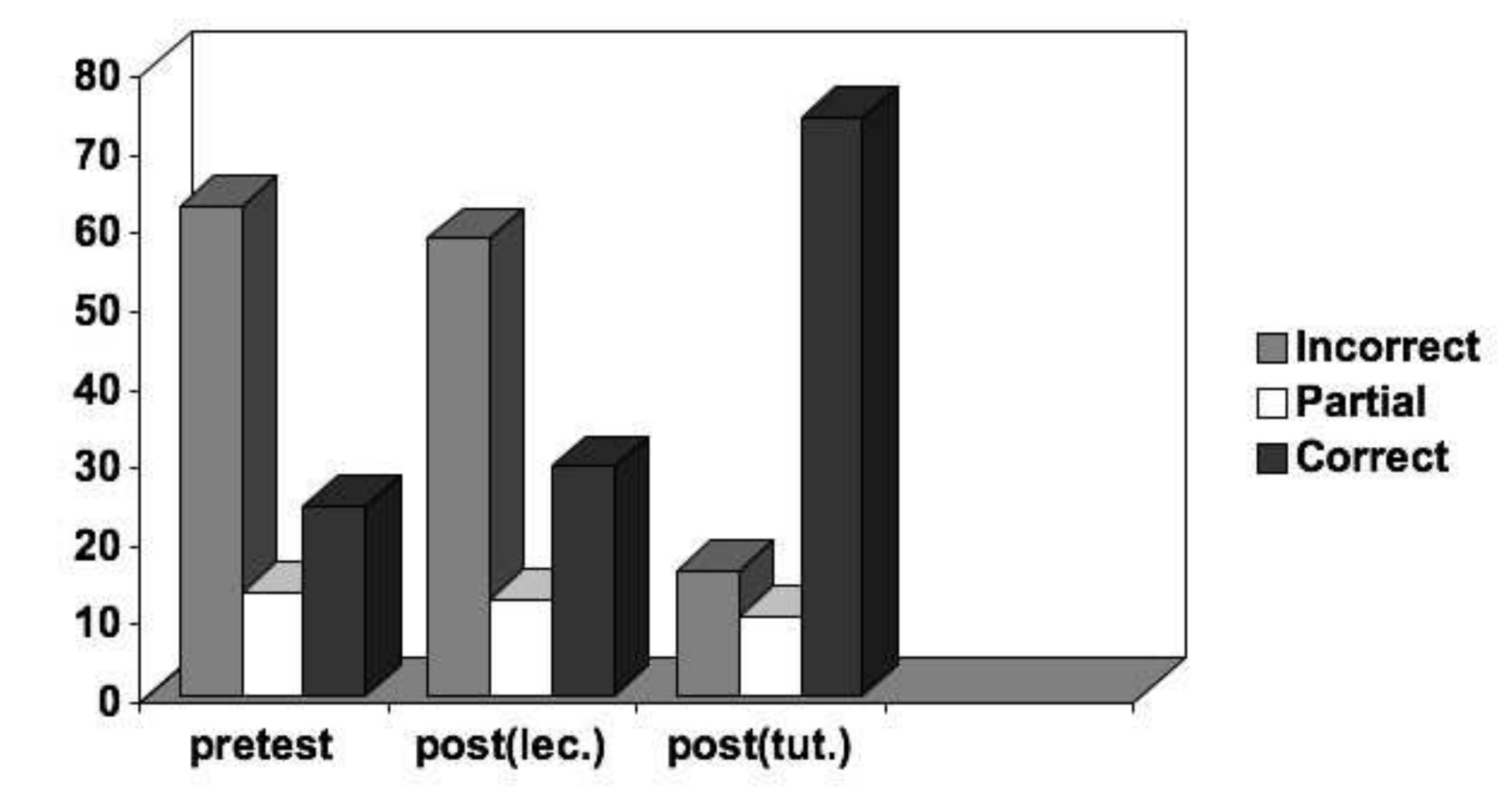}
\caption{
The average percentage of post-test responses that were answered incorrectly, partially correctly, or correctly.
}
\end{figure}

For example, 
when asked to show the distribution of the induced charges on a conducting shell with a point charge outside
the shell (similar to the multiple-choice question (1)), 
the students taught traditionally (via lecture only) continued to have trouble on the post-test. 
Compared to the traditionally taught students, $24\%$ more students who learned using the tutorials 
depicted the charge configuration correctly. 
Many students in the traditional group continued to draw induced charges on the inner walls of the conducting shell.

On the pre-/post-test free response questions that were similar to the multiple-choice question (2) with
the grounding wire, many traditionally taught students displayed difficulties similar to those on the pre-test. 
They still drew positive charges leaving the conductor via the grounding wire or shifted the positive charge downward assuming
that free electrons in the Earth will attract the positive charges on the conductor. Although some students
realized that the order in which the grounding wire and the charged comb are removed is important
in order to make the conductor develop a net charge via this process, they often did not provide an adequate
explanation for how this is the case. 
On the other hand, the tutorial group performed significantly better on the post-test questions about grounding. 
Not only could many of these students draw the final charge configuration on the conductor after induction and grounding processes, 
they drew and justified the proper charge configuration. 

\vspace*{-.22in}
\section{Conclusions}
\vspace*{-.10in}

We find that the concepts related to conductors and insulators covered in introductory physics are very challenging not only for 
the introductory physics students but also for the physics graduate students. 
Since homeworks or exams in upper-level courses seldom require students to make qualitative inferences from quantitative tools,
students do not develop a good grasp of the underlying concepts or learn to apply the formalism to physical situations.
The conceptual questions discussed here often require long chains of systematic reasoning. 
These difficulties motivated the development 
of tutorials about these concepts which were evaluated using a controlled study in which students were given
a pre-test and post-test before and after the traditional and tutorial instructions. The tutorials 
have been used by introductory students and physics majors and are helpful in improving students' understanding 
of these concepts.

\vspace*{-.15in}
\begin{theacknowledgments}
We are grateful to the NSF for award DUE-0442087.
\end{theacknowledgments}
\vspace*{-.15in}

\bibliographystyle{aipproc}

\end{document}